\documentclass[11pt]{iopart}
\usepackage{iopams}
\usepackage{epsfig}
\newcommand{\be}{\begin{equation}}
\newcommand{\ee}{\end{equation}}

\newcommand{\BE}{\begin{eqnarray}}
\newcommand{\EE}{\end{eqnarray}}

\newcommand{\erf}{{\rm erf}}

\newcommand{\bx}{\ensuremath{\mathbf{x}}}
\newcommand{\by}{\ensuremath{\mathbf{y}}}

\newcommand{\xt}{\widetilde{x}}
\newcommand{\yt}{\widetilde{y}}
\newcommand{\Th}[1]{\Theta(#1)}

\newcommand{\phit}{\widetilde{\phi}}

\newcommand{\boldpsi}{{\mbox{\boldmath $\psi$}}}

\newcommand{\avg}[1]{\left\langle{#1}\right\rangle}
\newcommand{\davg}[1]{\left\langle\left\langle{#1}\right\rangle\right\rangle}

\newcommand{\qt}{\ensuremath{\widetilde q}}
\begin{document}
\setlength{\parindent}{0em}
\title[Two-population replicator dynamics and number of Nash equilibria in matrix games]{Two-population replicator dynamics and number of Nash equilibria in random matrix games}

\author{Tobias Galla\dag\ddag
}

\address{\dag\ The Abdus Salam International Center for Theoretical Physics, Strada Costiera 11, 34014 Trieste, Italy}
\address{\ddag INFM-CNR, Trieste-SISSA Unit, V. Beirut 2-4, 34014 Trieste, Italy}

\begin{abstract}
We study the connection between the evolutionary replicator dynamics
and the number of Nash equilibria in large random bi-matrix
games. Using techniques of disordered systems theory we compute the
statistical properties of both, the fixed points of the dynamics and
the Nash equilibria. Except for the special case of zero-sum games one
finds a transition as a function of the so-called co-operation
pressure between a phase in which there is a unique stable fixed point
of the dynamics coinciding with a unique Nash equilibrium, and an
unstable phase in which there are exponentially many Nash equilibria
with statistical properties different from the stationary state of the
replicator equations. Our analytical results are confirmed by
numerical simulations of the replicator dynamics, and by explicit
enumeration of Nash equilibria.
\end{abstract}


\ead{\tt  galla@ictp.trieste.it}

\section{Introduction}

Replicator equations (RE) describe the evolution of populations of
species interacting through co-operation and competition. A fitness is
assigned to each species, dependent on the composition the population
and on the availability of resources, and species fitter than average
increase in concentration while the weight of species less fit than
average decreases in time. Replicator dynamics (RD) have found
widespread applications in game theory and economics as well as in
population dynamics where an equivalence to Lotka-Volterra equations
of theoretical biology can be established \cite{books}.

In the context of evolutionary game theory \cite{maynard} RE describe
games which are played repeatedly and where at any
time-step the interaction is between individuals randomly chosen out
of populations of agents.  Each player is here taken to follow a
pre-programmed strategy which they cannot change. Over time a payoff
is accrued, and agents reproduce as described
above, passing on their strategy to their descendents. This mechanism can be seen as  reinforcement learning from past experience \cite{farmer}.

Prior to the launch of this evolutionary approach to game theory the
analysis of games was mostly concerned with static aspects
\cite{nashneumann}, i.e. with strategically optimal actions by fully
rational players and the characterization of potential equilibrium
points. It is here assumed that the game is played only once.  Players
choose potentially stochastically from a set of available actions, and
payoffs are then paid to each participant dependent on the decisions
of all involved players. The key notion of a Nash equilibrium (NE)
then refers to a point in strategy space so that no player can
increase his payoff by unilaterally deviating from this point.

In this paper we extend existing work on single-population random
replicator dynamics \cite{opper} in which a transition between stable
and unstable regimes has been found as a function of a so-called
`co-operation pressure', and study the case of multi-population
models, i.e. interaction between {\em distinct} populations of
species. These correspond to so-called `asymmetric' games \cite{books}
in which there is more than one type of player, such as in the game
known as `the battle of the sexes' in which male and female types of
agents have different strategy sets at their disposal. In particular
we study the stability of two-population replicator systems, and the
relation to the number of NE in the corresponding bi-matrix games. We
here extend the work of Berg et al \cite{berg}, who computed the
number and statistics of NE of matrix games with random payoff
matrices in the absence of co-operation pressure and who identified an
exponentially large number of NE for all types of payoff correlations
considered, except for the case of zero-sum games in which there is a
unique NE. Our work thus extends the statistical mechanics approach to
replicator systems to the case of multiple populations, and
establishes a connection between the earlier works of
\cite{opper} and
\cite{berg}.

\section{Model definitions}
We will consider bi-matrix games between two players, one of type $X$
and one of type $Y$, with strategy sets $\Sigma_x$ and $\Sigma_y$, and
payoff functions $\mu_x,
\mu_y:\Sigma_x\times\Sigma_y\to\mathbb{R}$. We will here restrict to the case in which
$|\Sigma_x|=|\Sigma_y|=N$. Strategies available to players of type $X$
are labelled by $i\in\{1,\dots,N\}$, the ones available to $Y$ by
$j\in\{1,\dots,N\}$. The extension to more general cases with
different numbers of available strategies is straightforward.  Mixed
strategies are then probability distributions over $\Sigma_x$ and
$\Sigma_y$ respectively, described by vectors $\bx=(x_1,\dots,x_N)$
and $\by=(y_1,\dots,y_N)$, with $0\leq x_i\leq 1$ and $\sum_i x_i=1$,
and similarly for the $\{y_j\}$. Pure strategies are recovered as unit
vectors. If player $Y$ plays mixed strategy $\by$ then the expected
payoff for $X$'s pure strategy $i$ reads $\nu_i^x(\by)=\sum_j
\mu(i,j)y_j$. If both players play mixed strategies the expected
payoff for $X$ is denoted by $\nu_x(\bx,\by)=\sum_{i} x_i \nu_i^x(\by)$  and similarly for $\nu_y(\bx,\by)$.  A NE
is then a point $(\bx^\star,\by^\star)$ so that
$\nu_x(\bx^\star,\by^\star)=\mbox{max}_{\bx}\nu_x(\bx,\by^\star)$ and
$\nu_y(\bx^\star,\by^\star)=\mbox{max}_{\by}\nu_x(\bx^\star,\by)$, and
may be characterised by the following conditions
\BE
x_i^\star\left[\nu_i^x(\by^\star)-\nu_x(\bx^\star,\by^\star)\right]=0,&&\hspace{0.2cm}\nu_i^x(\by^\star)-\nu_x(\bx^\star,\by^\star)\leq 0, ~~~\forall i \label{eq:necondx}\\
y_j^\star\left[\nu_j^y(\bx^\star)-\nu_y(\bx^\star,\by^\star)\right]=0,&&\hspace{0.2cm}\nu_j^y(\bx^\star)-\nu_y(\bx^\star,\by^\star)\leq 0, ~~~\forall j. \label{eq:necondy}
\EE
The inequalities here ensure that no pure strategy delivers a higher payoff than obtainable at the Nash point, and the equalities enforce that all pure strategies which are actually played with non-zero probability yield the same payoff, and that all other pure strategies are not employed.

RE of evolutionary game theory assume large populations of $X$ and
$Y$-type players, respectively, with each individual playing a
pre-specified pure strategy. The replicators are hence the pure
strategies of the game under consideration and are copied without
error from parent to child. $x_i(t)$ denotes the relative concentration
of pure strategy $i$ in the $X$-population, and similarly for
$y_j(t)$. The replicator equations describing the evolution of the $X$
and $Y$ populations are then as follows
\BE\label{eq:dynxy}
\hspace{-2cm}\dot x_i(t)=x_i(t)\left[\nu_i^x(\by(t))-\kappa_x(t)\right], ~~~~\dot y_i(t)=y_i(t)\left[\nu_j^y(\bx(t))-\kappa_y(t)\right] 
\EE
where the $x_i(t)$ and $y_j(t)$ are now time-dependent variables, and
where $\kappa_x(t)=\sum_i x_i(t)\nu_i^x(\by(t))$, and similarly for
$\kappa_y(t)$ are the average payoffs of the $X$ and $Y$ type players
respectively. The replicator equations preserve the overall
normalisation $\sum_i x_i(t)=\sum_j y_j(t)=1$ in time. In terms of
population dynamics the weights $\{x_i(t)\}$ and $\{y_j(t)\}$ correspond to
relative concentrations of species, and the payoff functions $\nu_i^x$
an $\nu_j^y$ may be seen as their respective fitnesses. In
a biological setting RD thus describe the temporal evolution of
populations of species, where the concentrations of species fitter
than average grow in and where all other species decrease in relative
numbers. In the remainder of the paper we will use both the game
theoretical and the population dynamical language synonymously. Note that in the two-population system species of type $X$ interact only with species of type $Y$ and vice versa.

We will in the following be concerned with replicator systems where all
entries of the payoff matrices are Gaussian random
variables. Following Peschel and Mende \cite{books} as well as
\cite{opper} we will also introduce so-called `co-operation pressures'
$u_x\geq 0$ and $u_y\geq 0$, and will consider payoff functions of the
form
\BE
\hspace{-1.5cm}\nu_i^x(\bx,\by)=-2u_x x_i+\sum_{j} a_{ij} y_j, ~~~\nu_j^y(\bx,\by)=-2u_y y_j+\sum_{i} b_{ji} x_i.
\EE 
The role of $u_x,u_y$ will be clarified below. For $u_x=u_y=0$ one
recovers the cases studied in \cite{berg} (with payoff matrices
$a_{ij}=\mu_x(i,j),\,b_{ji}=\mu_y(i,j)$). Both the static and
dynamical properties of bi-matrix games are invariant under global
rescaling and shifts of all payoff matrix elements. Without
loss of generality we may therefore assume that the Gaussian variables
$a_{ij}$ and $b_{ji}$ are drawn from the following statistics
\BE
\overline{a_{ij}^2}=\overline{b_{ji}^2}=\frac{1}{N}, ~~~~ \overline{a_{ij}b_{kl}}=\delta_{il}\delta_{jk}\frac{\Gamma}{N}
\EE
(with $\overline{\cdots}$ an average over the distribution of
payoffs). The scaling with $N$ is here introduced to guarantee a well
defined thermodynamic limit. The parameter $\Gamma$ characterises the
correlations between the payoff matrices for the two different types
of players. If $\Gamma=-1$ then $a_{ij}=-b_{ji}$ so that the resulting
game is a zero-sum game (at vanishing co-operation pressures),
corresponding to a prey-predator relation in the population dynamical
setting. For $\Gamma=0$ one has uncorrelated payoff matrices, and in
the fully symmetric case $\Gamma=1$ the two interacting players always
receive equal payoff. Note also that for convenience we will re-scale
the $x_i$ and $y_j$ such that they obey the normalisations
$\sum_{i=1}^N x_{i}(t)=\sum_{j=1}^N y_j(t)=N$ at all times $t$. The
co-operation pressures $u_x$ and $u_y$ finally control the growth of
individual species, large values of $u_x$ drive the configuration
$\bx$ into the the interior of the simplex defined by $\sum_i x_i=N,
0\leq x_i\leq N$, and similarly for $u_y$, see the book by Peschel and
Mende for further details \cite{books}. Indeed, inspection of
Eqs. (\ref{eq:dynxy}) for $u_x, u_y\to\infty$ shows that the RD leads
to the fixed point $x_i=y_j\equiv 1$ for all species when started from
non-zero concentrations, corresponding to full co-operation and
maximal diversity. In a game theoretic setting large values of $u_x,
u_y$ favour the use of mixed strategies as opposed to pure strategies
located at the corners of the above simplices.

\section{Dynamics: generating functional analysis}

On here directly addresses the dynamics described by the replicator equations, and formulates an effective theory for macroscopic dynamical order parameters. The starting point of the analysis is the moment generating functional, defined as
\be
Z[\boldpsi^x,\boldpsi^y]=\davg{\exp\left(i\int dt\bigg\{\sum_{i=1}^N \psi^x_i(t) x_i(t)+\sum_{j=1}^N \psi^y_j(t) y_j(t)\bigg\}\right)},
\ee
where $\davg{\dots}$ denotes an average over trajectories
of the system, i.e. over solutions of the replicator equations
(\ref{eq:dynxy}).  As usual for disordered systems a closed, but
implicit set of equations describing the temporal evolution of a small
number of disorder-averaged macroscopic order parameters can be
derived. In the thermodynamic limit these observables turn out to be
given by the correlation functions $C_x(t,t'), C_y(t,t')$, the
response functions $G_x(t,t'), G_y(t,t')$ and the Lagrange parameters
$\kappa_x(t),\kappa_y(t)$. The correlation and
response matrices for the $X$-population are given by
\be
\hspace{-1.5cm}C_x(t,t')=\lim_{N\to\infty} N^{-1}\sum_{i=1}^N\overline{\davg{x_i(t)x_i(t')}}, ~~~~ G_x(t,t')=\lim_{N\to\infty} N^{-1}\sum_{i=1}^N\overline{\davg{\frac{\delta x_i(t)}{\delta \kappa_x(t')}}}, 
\ee 
and analogous definitions for $C_y, G_y$ apply.  These order
parameters are to be determined self-consistently as averages $C_x(t,t')=\avg{x(t)x(t')}_\star$, $G_x(t,t')=\avg{\delta x(t)/\delta
\kappa_x(t)}_\star$ (and analogously for $C_y,G_y$) over
realisations of the following pair of coupled stochastic effective processes
\BE
\dot x(t)&=&-x(t)\bigg[2u_x x(t)-\Gamma\int_{t_0}^t dt'~G_y(t,t')x(t')-\kappa_x(t)+\eta_x(t)\bigg] \\
\dot y(t)&=&-y(t)\bigg[2u_y y(t)-\Gamma\int_{t_0}^t dt'~G_x(t,t')y(t')-\kappa_y(t)+\eta_y(t)\bigg]
\EE

with $t_0$ the starting point of the dynamics.  The notation
$\avg{\dots}_\star$ refers to an average over the effective process,
i.e. over realisations of the noise variables $\{\eta_x(t)\}$ and
$\{\eta_y(t)\}$. The covariances of these noise variables are given by
$\avg{\eta_x(t)\eta_x(t'}_\star=C_y(t,t'),
\avg{\eta_y(t)\eta_y(t'}_\star=C_x(t,t')$ with no correlations between
$\eta_x$ and $\eta_y$. Finally, the Lagrange multipliers
$\{\kappa_x(t)\}$ and $\{\kappa_y(t)\}$ have to be chosen such that
the constraints $\avg{x(t)}_\star=1$ and $\avg{y(t)}_\star=1$ are
fulfilled at any time $t\geq t_0$.

Further progress can be made by assuming a fixed point of the
replicator equations, i.e. by inspecting time-independent solutions $x(t)\equiv x,\,y(t)\equiv y$ of
the effective processes (leading to constant correlation functions
$C_x(t,t')\equiv q_x, C_y(t,t')\equiv q_y$ and to stationary response
functions $G_x(\tau),G_y(\tau)$). Similarly to
\cite{opper} one derives the following equations characterising such
fixed point states 
\BE
\chi_x(2u_x-\Gamma\chi_y)=g_0(\Delta_x)&~~~& 
\chi_y(2u_y-\Gamma\chi_x)=g_0(\Delta_y)\nonumber \\
q_y^{-1/2}(2u_x-\Gamma\chi_y)=g_1(\Delta_x)&~~~&
q_x^{-1/2}(2u_y-\Gamma\chi_x)=g_1(\Delta_y)\nonumber\\
(q_x/q_y)(2u_x-\Gamma\chi_y)^2=g_2(\Delta_x)&~~~&
(q_y/q_x)(2u_y-\Gamma\chi_x)^2=g_2(\Delta_y)\label{eq:fp}
\EE
with $\Delta_x=\kappa_x/\sqrt{q_y}$, $\Delta_y=\kappa_y/\sqrt{q_x}$
and $g_n(\Delta)=\int_{-\infty}^\Delta
dz\frac{e^{-z^2/2}}{\sqrt{2\pi}}(\Delta-z)^n$ for $n\in\{0,1,2\}$. $\chi_x$
is here the integrated response $\chi_x=\int dt G_x(\tau)$, and
similarly for $\chi_y$. A linear stability analysis shows that such
fixed points become unstable at the point at
which $(\chi_x\chi_y)^2=\phi_x\phi_y$, leading to an unstable and
non-ergodic phase at low co-operation pressures. For $u_x=u_y$, which
we will mostly consider in the following, one finds
$\chi_x=\chi_y$ and $q_x=q_y$, and Eqs. (\ref{eq:fp}) as well as the
stability condition reduce to those of a single-population replicator
system studied in
\cite{opper}. In the unstable phase solutions of (\ref{eq:fp}) can no longer be expected to describe the stationary states of the RE accurately.
\section{Statics: replica analysis}
The starting point of the replica analysis of the statics of the model
is the observation that for symmetric couplings, i.e. $a_{ij}=b_{ji}$
($\Gamma=1$) the replicator equations (\ref{eq:dynxy}) can be written
in the form $\dot x_i=x_i\left(\partial_{x_i} {\cal
H}(\bx,\by)-\kappa_x(t)\right)$ and similarly for $\dot y_j$, with
${\cal
H}(\bx,\by)=\frac{1}{2}\sum_{i,j=1}^N\left[x_iy_j(a_{ij}+b_{ji})\right]-u_x\sum_{i=1}^N
x_i^2-u_y\sum_{j=1}^N y_j^2$, so that the stationary states of the RD
at $\Gamma=1$ correspond to extrema of ${\cal H}$. The computation of
these is straightforward and based on the evaluation of $-\beta
f=\lim_{N\to\infty}N^{-1}\overline{\ln Z(\beta)}$. $f$ is here the
disorder-averaged free energy density at temperature
$T=\beta^{-1}$. $Z(\beta)$ stands for the partition function
corresponding to ${\cal H}$:
\BE
\hspace{-2cm}Z(\beta)&=&\left[\prod_{i=1}^N\int_0^\infty dx_i\prod_{j=1}^N\int_0^\infty dy_j\right] \delta\left(\sum_{j-1}^N x_j-N\right) \delta\left(\sum_{j-1}^N y_j-N\right)e^{-\beta{\cal H}(\bx,\by)}\label{eq:zet}.
\EE 
We will not report the details of the straightforward replica calculation, but will only note that a replica-symmetric ansatz leads to a set of equations identical to (\ref{eq:fp}) at zero temperature and in the thermodynamic limit (with $\Gamma=1$). For $\Gamma<1$ no Lyapunov function ${\cal H}$ can be found and the replica approach is inapplicable.

\section{Annealed calculation of the number of Nash equilibria}
Finally, the number and statistics of the NE of the corresponding bi-matrix game may be computed by direct integration over phase space enforcing conditions (\ref{eq:necondx},\ref{eq:necondy}) through suitable delta- and step-functions. It is here convenient to set $\xt_i=x_i$ if $x_i>0$ and $\xt_i=-2u_xx_i+\sum_j a_{ij}y_j-\kappa_x$ if $x_i=0$, and similarly for $\yt_j$ \cite{berg}. The above conditions (\ref{eq:necondx}) may then be written as 
\be
I_i^x({\bf \xt},{\bf \yt})\equiv\xt_i\Th{-\xt_i}-\left(-2u\xt_i\Th{\xt_i}+\sum_j a_{ij}\yt_j\Th{\yt_j}-\kappa_x\right)=0
\ee
and (\ref{eq:necondy}) translates into $I_j^y({\bf \xt},{\bf \yt})=0$
with an analogous expression $I_j^y({\bf \xt},{\bf
\yt})$. $\Th{\cdot}$ is the step-function.  The number of NE at
payoffs $\kappa_x$ and $\kappa_y$ is then given by
\BE
{\cal N}(\kappa_x,\kappa_y)&=&\int D{\bf \xt} D{\bf \yt} \delta\left(\sum_i \xt_i\Th{\xt_i}-N\right)\delta\left(\sum_j \yt_i\Th{\yt_i}-N\right)\nonumber \\
&&\times \prod_i\delta(I_i^x({\bf \xt},{\bf \yt}))\prod_j\delta(I_j^y({\bf \xt},{\bf \yt}))~~|\det D|.
\EE
$\det D$ is a normalising determinant. Performing the disorder-average
in an annealed approximation, one converts the computation into a
saddle-point problem in the thermodynamic limit. We set $u_x=u_y=u$
for simplicity and find exponential domination at equal payoff
$\kappa_x=\kappa_y=\kappa$ and $\lim_{N\to\infty}\frac{1}{N}\ln
\overline{{\cal N}(\kappa)}=S(\kappa)$, where
\BE\label{eq:entr}
\hspace{-2cm}S(\kappa)=\mbox{extr}_{\{E,R,q,\qt,\phi,\phit\}}\Bigg\{2E-\Gamma R^2+2\qt q+2\phit\phi+2\ln\Bigg[H\left(-\frac{\kappa}{\sqrt{q}}\right)+\frac{e^{-\phit}}{\sqrt{(2u-\Gamma R)^2+2q\qt}}~~~~~~~~~\nonumber \\ \hspace{-1.5cm} \times H\left(\frac{(2u-\Gamma R)\kappa/q+E}{\sqrt{q^{-1}(2u-\Gamma R)^2+2\qt}}\right)
\times
\exp\left(-\frac{\kappa^2}{2q}+\frac{(E+(2u-\Gamma R)\kappa/q)^2}{4\qt+2(2u-\Gamma R)^2/q}\right)\Bigg]+g\Bigg\}
\EE
We have here set abbreviated
$H(x)=\frac{1}{2}\left(1-\erf(x/\sqrt{2})\right)$, and $g$ denotes the
contribution from the normalising determinant, which we do not report
explicitly. For $u=0$ we recover the result of
\cite{berg}. Extremisation of (\ref{eq:entr}) leads to an annealed
upper bound of the logarithmic number of NE at payoff $\kappa$.
\section{Results}

\begin{figure}[t]
\vspace*{1mm}
\begin{tabular}{cc}
\epsfxsize=60mm  \epsffile{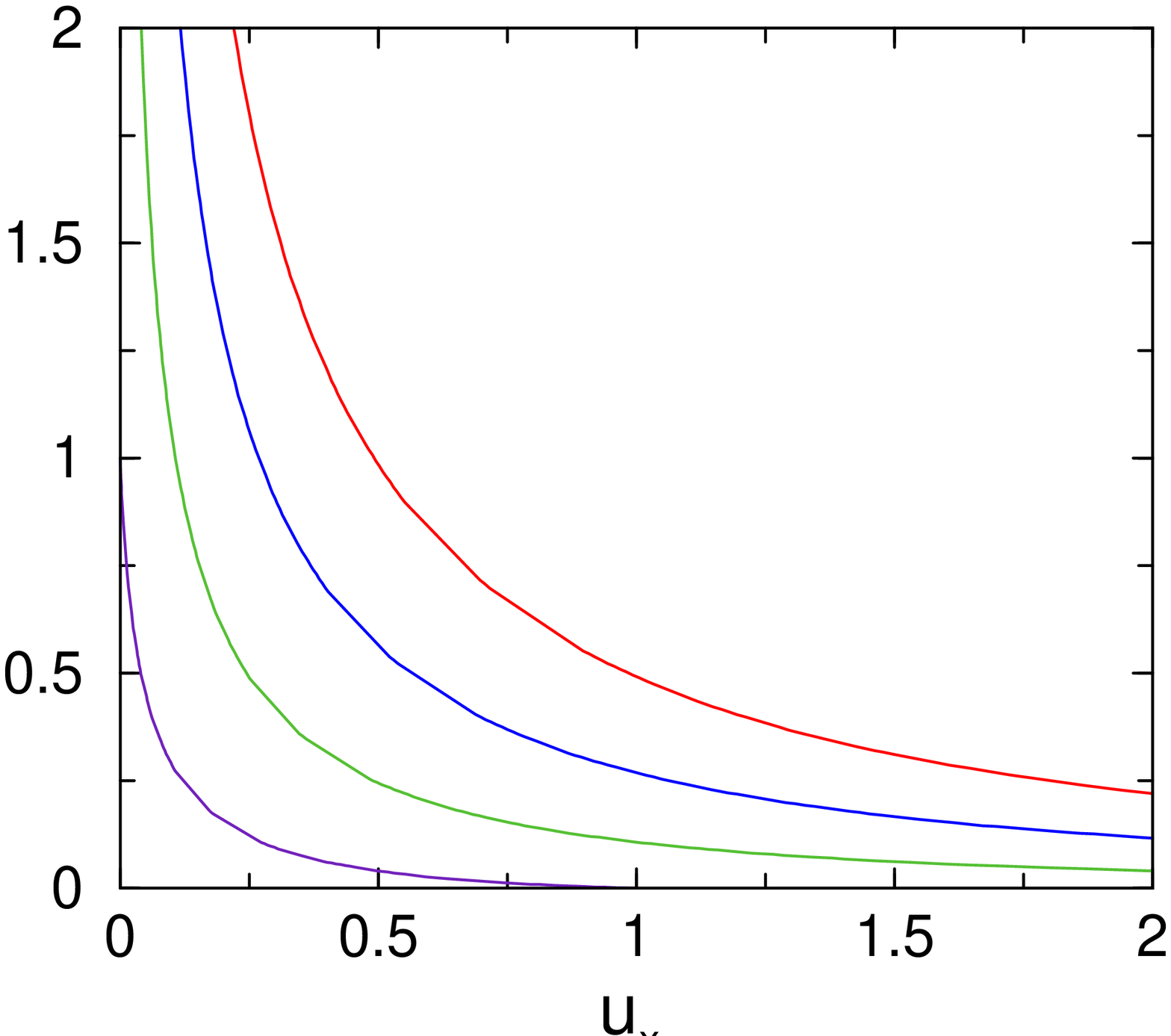} ~~~&~~~
\epsfxsize=60mm  \epsffile{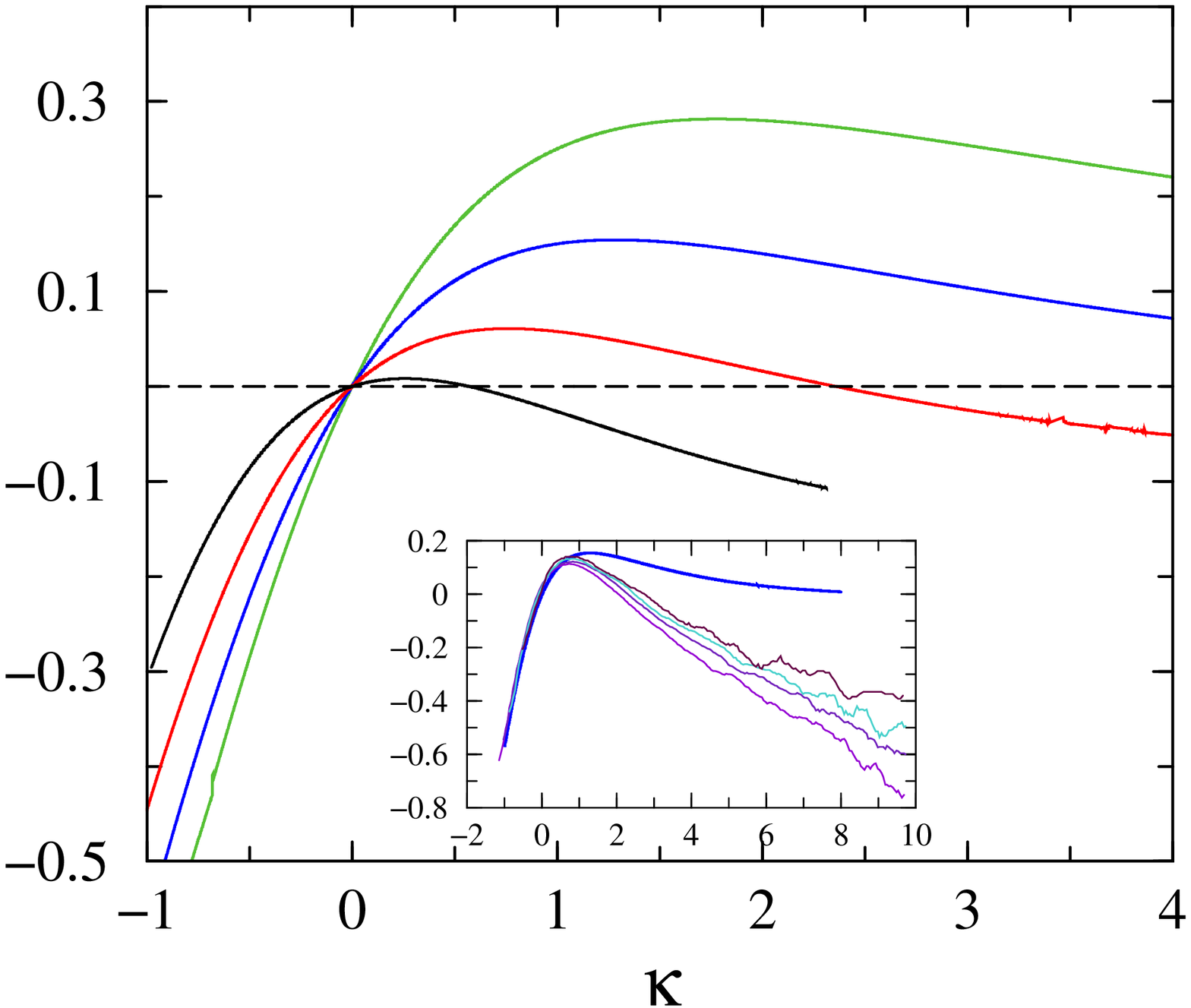}
\end{tabular}
\vspace*{4mm} \caption{\\{\bf Left:} (Colour on-line) Phase diagram of the two-population replicator system. Lines separate the stable phase (up-right) from the instable one, and are plotted for $\Gamma=1,0.5,0,-0.5$ from top to bottom.\\ {\bf Right:} (Colour on-line) Main panel: entropy of number of Nash equilibria for the uncorrelated game $\Gamma=0$ at $u=0, 0.1, 0.2, 0.3$ (from top to bottom at the maximum). Inset: comparison with numerical data from support enumeration ($u=0.1$). Upper curve is the theoretical line, others are from support enumeration of single-population models at $N=14, 16, 18, 20$ (bottom to top). A running average has been performed to smoothen numerical data.}\label{fig1}
\end{figure}

\begin{figure}[t]
\vspace*{1mm}
\begin{tabular}{cc}
\epsfxsize=60mm  \epsffile{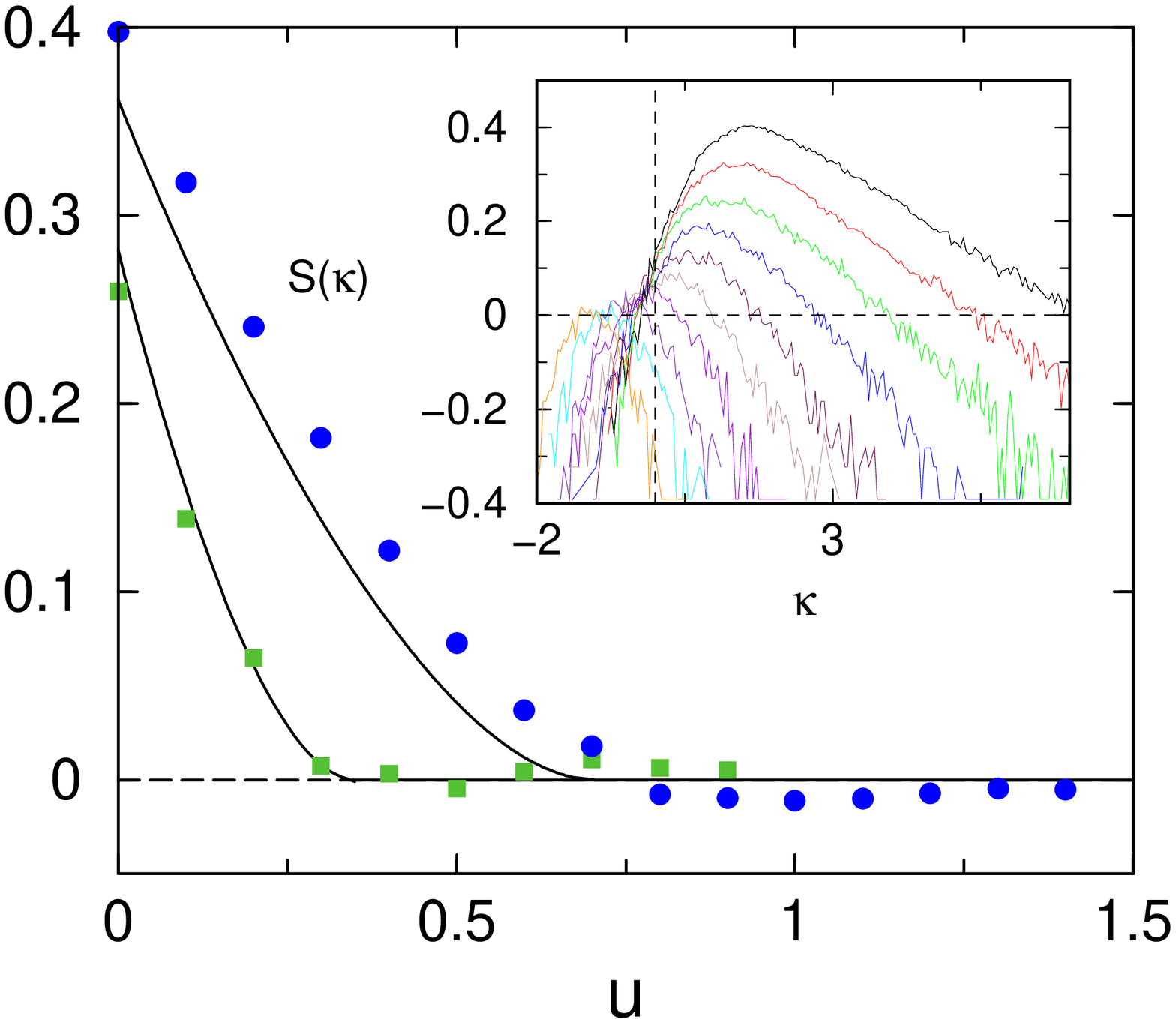} ~~&~~
\epsfxsize=60mm  \epsffile{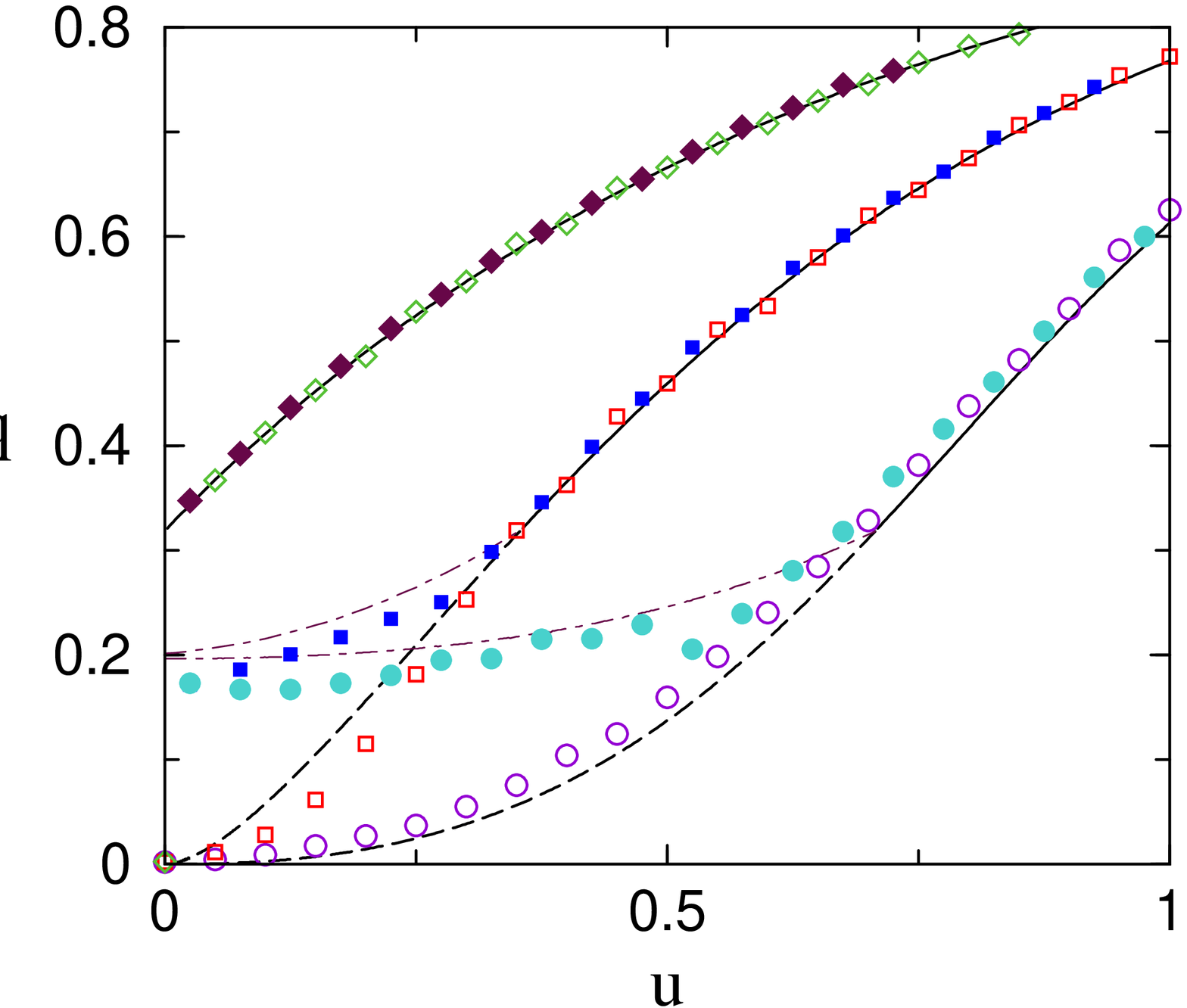}
\end{tabular}
\vspace*{4mm} \caption{\\ {\bf Left:} (Colour on-line) Entropy $S_m$ of Nash equilibria as a function of $u$, for $\Gamma=0$ (left curve) and $\Gamma=1$ (right curve). Solid lines from theory, symbols from explicit enumeration of single-population systems with $N=20$, averaged over $1000$ samples. Inset: Raw data for $S(\kappa)$ at $\Gamma=1$ for $u=0.9, 0.8, \dots, 0.0$ from left to right.\\ {\bf Right:} (Colour on-line) Diversity parameter $q^{-1}$ at $\Gamma=-1,0,1$ (from left to right). Solutions of (\ref{eq:fp}) depicted as solid lines where replicator equations have stable fixed points, continued into the instable phase as dashed lines. Open symbols are results from simulations of bi-replicator systems with $2N=1000$ species. Dashed-dotted curves show theory for Nash equilibria for $u\leq u_c(\Gamma)$, open symbols are numerical results for typical NE as obtained from iterated Lemke-Howson algorithm applied to bi-matrix systems of size $2N=100$, averaged over $200$ samples. }
\label{fig2}
\end{figure}

Our results are summarised by figures \ref{fig1} and
\ref{fig2}. Fig. \ref{fig1}a shows the phase diagram of the
bi-population replicator model at different symmetry parameters
$\Gamma$. For $\Gamma>-1$ one finds a transition line, with unique
fixed points of the RE at large co-operation pressures to the
top-right, and an unstable regime below. For $\Gamma=-1$ no transition
is observed, in this case there is one unique fixed-point of the
dynamics for any $(u_x,u_y)$. In the remaining figures we restrict to
the case $u_x=u_y=u$. The transition between the stable
and unstable phases then occurs at co-operation pressure
$u_c(\Gamma)=(1+\Gamma)/(2\sqrt{2})$. In Fig. \ref{fig1}b we depict
typical curves $S(\kappa)$ of the entropy of NE at payoff $\kappa$ as
obtained from Eq. (\ref{eq:entr}). These curves typically show a
maximum $S_m(u,\Gamma)$ at intermediate values of the payoff $\kappa$,
indicating the number of dominating NE. If $S_m>0$ a large number of
Nash equilibria is present in the thermodynamic limit, while for
$S_m<0$ NE are exponentially suppressed. If $S_m=0$ a single NE
prevails \cite{berg}. The inset of Fig. \ref{fig1}b shows a comparison
between the analytical results for $S(\kappa)$ at a fixed combination
of $u$ and $\Gamma$ with numerical data from an explicit enumeration
of NE. We here use a direct support enumeration scheme, which we apply
to an equivalent single-population game in order to reduce the
required computing time. Accessible system sizes are limited to
$N\approx 20$, and while results are consistent with the theory, finite
size effects can be significant.

Fig. \ref{fig2}a shows the entropy $S_m$ of the dominating
NE as a function of $u$ at fixed values of the symmetry parameter
$\Gamma$. One finds a large number of NE ($S_m>0$) for
$u<u_c(\Gamma)$, and a unique NE above $u_c$ ($S_m=0$). At $u_c$ the
spectra $S(\kappa)$ take their maxima at zero payoff $\kappa=0$. The
transition point $u_c(\Gamma)$ coincides with the one separating the
stable and unstable phases of the RD. Results from an explicit
enumeration of NE are consistent with this transition although direct
quantitative comparison between simulations and theory appears
difficult due to finite size effects.  An example of raw data obtained
from enumeration of NE is shown in the inset of
Fig. \ref{fig2}a.

The saddle point extremisation of (\ref{eq:entr}) allows one also to
compute the statistics of the dominating NE (in the annealed
approximation) and to compare with the stationary states of the RE. We
here focus on the order parameter $q^{-1}$ which serves as a measure
of the diversity of the eco-system \cite{simpson}, with large values
of $q^{-1}$ corresponding to many surviving species (equivalently many
pure strategies played with non-zero probability).  Measurements of
$q^{-1}$ in the stationary states of the RE and numerical results for
the NE are shown in Fig. \ref{fig2}b. The latter are here obtained
using a repeated Lemke-Howson algorithm \cite{lemke}. Note that the
fixed-point theory of the RE applies only above $u_c$ but has been
continued below as dashed lines. Quantitative deviations between
numerical and theoretical results for NE are due to finite-size or
sampling effects and the annealed approximation of the theory. The
analytical and numerical results show that the diversity of species in
the NE coincides with that at the fixed points of the RD above
$u_c(\Gamma)$, but that they differ from each other below the
transition. Very similar results are found for the number of surviving
species. We note that conditions
(\ref{eq:necondx},\ref{eq:necondy}) are necessary but not sufficient
for the stability of fixed points of the RE so that stable fixed
points are always NE but not necessarily vice versa. Results thus
indicate that NE are statistically distinct from potential stable
attractors of the dynamics below $u_c$ (and more numerous). Note also
that below $u_c$ marginally stable fixed points are suppressed for
$\Gamma<1$ \cite{opper} and hence volatile, possibly chaotic behaviour
is observed in simulations. For $\Gamma=1$ the dynamics converges to a
fixed point also below $u_c$, but is sensitive to initial
conditions (see also \cite{opper,galla}).

\section{Summary and concluding remarks}

Our results may be summarised as follows: for games different from the
zero-sum type (i.e. for any $\Gamma>-1$) one finds a dynamical
instability of the fixed points of the two-population replicator
equations at $u_c(\Gamma)=(1+\Gamma)/(2\sqrt{2})$. For $\Gamma=1$ this
instability coincides with an AT-instability of the replica symmetric
solution of the statics \cite{parisi}. Above $u_c$ all three
approaches (dynamics, NE and replica theory where applicable) lead to
the same order parameters, describing an ergodic
state, in which there is one unique NE which coincides with the unique
fixed point of the RD (and with the unique extremum of ${\cal H}$ for
$\Gamma=1$). Below $u_c$ ergodicity is broken, the stationary state of
the RD is no longer unique, and there is an exponential number of
NE. The statistics of the NE differs from those of the stationary
states of the dynamics. The observations generalise the results of
\cite{berg}, which were concerned with the case of vanishing
co-operation pressure $u$. At $u=0$ the RD is instable for all
$\Gamma>-1$, and marginally stable for zero-sum games,
$\Gamma=-1$. Except for the latter case the system is therefore in the
non-ergodic phase of the RE, in line with the reported exponential
number of NE in \cite{berg}. At $\Gamma=-1$ one is precisely at the
phase transition at $u_c(\Gamma=-1)=0$ and finds a zero-entropy of dominating NE.

In conclusion we have investigated bi-matrix games and two-population
replicator systems with tools of disordered systems theory, and have
have computed the number and typical properties of NE and
of the fixed points of the corresponding replicator equations. Several
extensions of the present work can be considered, including cases of
correlations between different rows and/or columns of the payoff
matrices, populations with different numbers of strategies at hand and
systems with more than two types of players, as well as the study of other stability concepts of evolutionary game theory in the context of random replicator systems.

\section*{Acknowledgements}
This work was supported by the European Community's Human Potential
Programme under contract HPRN-CT-2002-00319, STIPCO. The author would
like to acknowledge discussions with Manfred Opper, and would like to
thank Johannes Berg for making available his software routines as well
as details of calculations reported in \cite{berg}.

\end{document}